\documentclass[conference]{IEEEtran}
\IEEEoverridecommandlockouts
\usepackage{cite}
\usepackage{amsmath,amssymb,amsfonts}
\usepackage{algorithmic}
\usepackage{graphicx}
\usepackage{textcomp}
\usepackage{enumitem}
\usepackage{graphicx}
\usepackage{hyperref}
\usepackage{caption}
\usepackage{booktabs}
\usepackage{tabularx}
\usepackage{diagbox} 
\usepackage{tcolorbox}
\tcbuselibrary{skins}
\usepackage[table]{xcolor}
\usepackage{array}
\usepackage{makecell}

\usepackage{diagbox}
\usepackage{multirow}
\usepackage{colortbl} 
\definecolor{Gray}{gray}{0.92}
\usepackage[table]{xcolor}
\definecolor{LightCyan}{rgb}{0.88,1,1}
\arrayrulecolor{black}
\usepackage{hhline}

\usepackage{multirow}
\usepackage[flushleft]{threeparttable}
\usepackage{tablefootnote}
\renewcommand\IEEEkeywordsname{Index Terms}
\def\BibTeX{{\rm B\kern-.05em{\sc i\kern-.025em b}\kern-.08em
    T\kern-.1667em\lower.7ex\hbox{E}\kern-.125emX}}

\graphicspath{{Images/}}

\begin{document}


\title{Method-level Change-proneness: A Better Metric for Black-box Test Suite Minimization}

\author{\IEEEauthorblockN{Md. Siam}
\IEEEauthorblockA{\textit{Institute of Information Technology} \\
\textit{University of Dhaka}\\
Dhaka, Bangladesh \\
\texttt{bsse1104@iit.du.ac.bd}}
\and
\IEEEauthorblockN{Kazi Sakib}
\IEEEauthorblockA{\textit{Institute of Information Technology} \\
\textit{University of Dhaka}\\
Dhaka, Bangladesh \\
\texttt{sakib@iit.du.ac.bd}}
}

\maketitle

\begin{abstract}

    Test Suite Minimization (TSM) reduces the size of test suites while preserving their fault detection capability. In black-box TSM, reduction is performed without analyzing production code. While several black-box TSM approaches have explored metrics like test logs or test similarity, those often suffer from scalability and efficiency issues. On the other hand, change-proneness (CP), recently emerged as an efficient and scalable alternative metric, has only been applied at class level. To accurately identify fault-revealing test cases, we propose CP at finer-grained method-level and implement Method-level Change-proneness based Test-suite Minimization (\textit{MCTM}). \textit{MCTM} first calculates CP for each method from version control metadata, then determines the dependency between test cases and methods by analyzing the test-code call-graph. Next, it scores the association between test cases and their invoked methods using statistical measures such as Average, Geometric Mean etc. Finally, test cases with the highest scores are selected to form the reduced suite. Evaluation on 15 open-source Java projects with 635 buggy versions shows \textit{MCTM} achieves 0.93 accuracy and 0.94 fault detection rate on average, significantly outperforming class-level CP and similarity-based approaches while maintaining superior efficiency.
  
\end{abstract}


\section{Introduction}

Test Suite Minimization (TSM) aims to reduce the number of test cases while preserving the test suite’s ability to detect faults \cite{DBLP:journals/stvr/YooH12, DBLP:journals/access/KhanLJA18}. When TSM is performed without utilizing production code, it is referred to as black-box\cite{DBLP:conf/icse/PhilipBKMN19,DBLP:conf/icse/PanGB23}. Various metrics such as test logs and test code similarity have been explored to estimate a test case’s fault detection potential, but these metrics often suffer from poor scalability, inefficiency in terms of execution time, or high resource demands \cite{10697930}. Recently, change-proneness (CP) has been explored as a more scalable and efficient metric alternative. \cite{siam2025exploratory}. However, the approach leveraging CP operates on class-level \cite{siam2025exploratory}, which is coarse-grained and may overlook important details such as the specific impact of changes within individual methods which is essential for effective TSM\cite{hata2010reconstructing, DBLP:conf/icse/HataMK12}.

Prior studies in software maintenance and defect prediction demonstrate the benefits of analyzing software artifacts at finer granularity. Metrics computed at the method level capture localized modifications and context-specific characteristics that are strongly associated with defects such as complexity, coupling, and churn \cite{catal2009investigating}. Compared to class or file-level metrics, method-level features have been shown to achieve higher accuracy in identifying fault-prone components and localizing buggy regions \cite{DBLP:conf/msr/GigerPG11, hata2010reconstructing}. Such fine-grained analysis also enables robust performance on large systems and enhances test case prioritization in evolving codebases \cite{luo2010software, mahdieh2022test}. These findings suggest that extending CP analysis to the method level could provide more precise guidance for selecting test cases in black-box TSM. Achieving this requires quantifying CP at the method level and scoring test cases based on their association with change-prone methods. To be viable for black-box TSM, the approach must be effective in fault detection while remaining efficient and lightweight.

Several approaches have been proposed for TSM in black-box. Philip et al.\cite{DBLP:conf/icse/PhilipBKMN19} introduced a black-box technique that leverages test code, commit complexity, historical test logs and version control (VC) metadata to train classification models for reducing test suites. Cruciani et al.\cite{DBLP:conf/icse/CrucianiMVB19} proposed a clustering-based method using test code diversity as the core metric. Building on this, Pan et al. developed AST-based (\textit{ATM})\cite{DBLP:conf/icse/PanGB23} and Large Language Model (LLM) - based (\textit{LTM})\cite{10697930} approaches, both relying on test case similarity and evolutionary search. Most recently, Siam et al.\cite{siam2025exploratory} introduced CP as a lightweight alternative metric for black-box TSM. Their approach leverages VC metadata to calculate CP values at the class level. These values are then used to score and prioritize test cases. While this work demonstrated the effectiveness of CP-based test selection, it operated exclusively at class-level granularity, which may miss fine-grained change patterns that occur at the method level within classes. To the best of our knowledge, no prior work has investigated the application of CP at method-level granularity for black-box TSM.

In this study, we investigate method-level CP as a metric for black-box TSM and integrate it into an approach called \textit{MCTM} (Method-level Change-proneness based Test-suite Minimization). \textit{MCTM} begins by extracting change metrics for each method from the project’s Git history. For each buggy version, method-level changes are analyzed from project inception to the current version. CP is then calculated using two metrics: Change Frequency (\textit{ChgFrq}), which measures how often a method changes, and Change Extent (\textit{ChgExt}), which captures the magnitude of those changes \cite{DBLP:conf/ease/ArvanitouACA17}. Once high-CP methods are identified, \textit{MCTM} determines the association of the test-cases with the change-prone methods. To do this, it builds a call graph of the test code and analyzes it to map test cases to the methods they invoke. Each test case is scored based on its association with change-prone methods using various statistical measures, including Average, Geometric Mean  etc. Finally, test cases with the highest association scores are selected to form the reduced test suite.

We evaluate MCTM on 15 Java projects comprising 635 buggy versions, testing 8 configurations across three minimization budgets (retaining 25\%, 50\%, and 75\% of the original test suite). Our evaluation addresses the following research questions:

\begin{itemize}
    \item RQ1: \textit{How effective is method-level CP in guiding black-box TSM?}  
    At a 50\% minimization budget, \textit{MCTM} demonstrated strong effectiveness, achieving an average fault detection rate of 0.94 and accuracy of 0.93, with the best results obtained using the Change Frequency metric and Average aggregation. It also maintained high efficiency, requiring only 0.98 minutes per version across all configurations. Results were consistent for other minimization budgets (25\% and 75\%).

    \item RQ2: \textit{How does method-level CP perform compared to class-level in terms of effectiveness and efficiency? } 
    The best configuration of \textit{MCTM} significantly outperformed the best configuration of the class-level CP-based approach, \textit{CTM}, in fault detection capability. It achieved +0.23 accuracy gain and +0.19 FDR gain across all projects and budgets. While introducing finer granularity, \textit{MCTM} preserved practical efficiency, running with only a modest overhead (mean runtime of 0.98 minutes vs 0.44 minutes per version). 

    \item RQ3: \textit{How does \textit{MCTM} perform against similarity-based approaches (ATM and LTM)?}  
    Compared to similarity-based methods, \textit{MCTM} achieved significant gains in both effectiveness and efficiency. It improved accuracy by 0.26 over ATM and 0.22 over LTM, with an FDR improvement of 0.10 over LTM, while reducing execution time by 2.65x. 
\end{itemize}

These findings demonstrate that method-level CP enables more effective black-box TSM than existing class-level CP and similarity-based approaches while being practically efficient.

\section{Related Work}

Test Suite Minimization (TSM) focuses on improving the efficiency of software testing by removing redundant test cases from a test suite while preserving its fault detection capability \cite{DBLP:journals/stvr/YooH12, DBLP:journals/ijinfoman/KhanLAAC16, DBLP:journals/access/KhanLJA18}. By reducing the size of test suites, TSM helps lower testing time and resource consumption without compromising the quality of testing outcomes.

Extensive research has been conducted on TSM, leading to a variety of techniques in the literature \cite{DBLP:journals/stvr/YooH12, DBLP:journals/access/KhanLJA18, DBLP:conf/icse/CrucianiMVB19}. Broadly, these techniques can be categorized into two categories: white-box and black-box approaches. White-box TSM relies on production code, code coverage data, or requirement models to guide test case selection\cite{DBLP:journals/access/KhanLJA18}. In contrast, black-box TSM avoids dependence on production code and instead leverages information available from test code, version control metadata, or commit history to identify valuable test cases\cite{DBLP:conf/icse/CrucianiMVB19, DBLP:conf/icse/PhilipBKMN19}.

Existing test suite minimization (TSM) techniques predominantly employ white-box strategies. These include (a) greedy heuristics, which iteratively select test cases based on their coverage of code elements like statements or branches \cite{miranda2017scope, noemmer2019evaluation, mohanty2020ant}, (b) search-based techniques, which use genetic algorithms or other evolutionary strategies to find optimal subsets of test cases that balance coverage and size \cite{hemmati2013achieving, zhang2019uncertainty}, (c) clustering-based methods, which group test cases by the similarity of their coverage profiles and select representatives from each cluster \cite{liu2011user, coviello2018clustering}, (d) model-based approaches, which analyze system models such as UML diagrams or state machines to evaluate test case relevance\cite{arrieta2019pareto}, and (e) requirement-driven strategies\cite{lin2014test, lin2012reducing}, which prioritize test cases based on their traceability to functional or non-functional requirements.

Although these methods have shown effectiveness in controlled experiments, their adoption in practice faces significant barriers. In industrial settings, production codebases are often very large and complex, making the analysis and instrumentation required for white-box TSM impractical at scale\cite{DBLP:conf/icse/CrucianiMVB19}. Additionally, the process of collecting and analyzing code coverage data can introduce considerable overhead, with studies reporting up to 30\% additional testing time \cite{DBLP:conf/icse/PanGB23, DBLP:conf/icse/CrucianiMVB19}. Such constraints limit the scalability and practical applicability of white-box TSM in fast-paced development environments.

To address these challenges, researchers have explored black-box TSM, which eliminates the dependency on production code by utilizing only test code or external metadata. Black-box approaches have gained attention for their ease of integration and lower computational overhead \cite{DBLP:conf/icse/PanGB23, DBLP:conf/icse/CrucianiMVB19, siam2025exploratory, 10697930}. Philip et al. \cite{DBLP:conf/icse/PhilipBKMN19} proposed the first black-box TSM approach, FastLane, using commit risk, version control metadata and test logs in a logistic regression model. They leverage the regression model to determine whether to run or skip the execution of a test case. FastLane achieved an 18.04\% reduction in test suite size while retaining 99.99\% of faults. However, it depends on historical test data, which is not always available \cite{siam2025exploratory}.

Cruciani et al.\cite{DBLP:conf/icse/CrucianiMVB19} introduced FAST-R, a black-box approach that relies solely on the source code of test cases. FAST-R transforms test code into vectors using a term frequency model \cite{johnson1984extensions} and applies random projection \cite{turney2010frequency} to reduce their dimensionality. Clustering algorithms are then used on these vectors, with cluster centroids selected to form the minimized test suite. Their results showed that FAST-R was significantly more efficient than white-box techniques in terms of minimization time. However, although FAST-R performed well for C projects, it showed noticeably lower fault detection capability in Java. This makes it less suitable for such contexts.

To address this limitation, Pan et al. \cite{DBLP:conf/icse/PanGB23} proposed \textit{ATM}, a TSM approach based on syntactic similarity. They removed irrelevant code elements, i.e. comments and test oracles, transformed the remaining code into Abstract Syntax Trees (ASTs). Then they computed pairwise similarity of ASTs to guide test selection using evolutionary search techniques (GA and NSGA-II). While effective in fault detection, \textit{ATM} struggled with scalability. To overcome this, \textit{LTM} \cite{10697930} was introduced, which captures semantic similarity using large language models such as UniXcoder and CodeBERT. It generates vector embeddings for test cases and computes similarity using cosine similarity and Euclidean distance. Although \textit{LTM} improved scalability, it requires GPU support for efficient execution.

Siam et al. \cite{siam2025exploratory} addressed this issue by introducing change proneness (CP) as a metric for TSM. Their black-box approach leverages version control metadata to identify change-prone classes and measures each test case’s association with these classes. The technique demonstrated significant efficiency gains over \textit{ATM} while maintaining comparable fault detection capability. However, analyzing change proneness at the class level remains coarse-grained and may miss critical details such as the distribution and intensity of changes within classes, which are important for effective test suite reduction \cite{hata2010reconstructing, DBLP:conf/icse/HataMK12}.

Prior work highlights the advantages of analyzing software artifacts at finer granularities, with method-level change metrics often outperforming coarser file or class-level metrics in fault prediction and localization \cite{DBLP:journals/corr/abs-2502-17908}. Method-level metrics capture localized modifications and contextual information, enabling more accurate bug predictions and effective fault localization \cite{DBLP:conf/msr/GigerPG11, sohn2019empirical}. Their effectiveness largely stems from the ability to represent factors such as complexity, coupling, and churn, which are strongly linked to defects \cite{catal2009investigating}. Luo et al.\cite{luo2010software} demonstrated that models leveraging over 40 method-level metrics perform robustly even on large systems. More recently, combining fault-proneness estimations with method-level change data has been shown to enhance test case prioritization\cite{mahdieh2022test, wen2018well}. Change metrics, when calculated at method-level, have been proven effective for bug identification\cite{hata2010reconstructing, DBLP:journals/ijsinnov/AbbasA22, DBLP:journals/corr/abs-2502-17908}. 

These suggest that CP at method-level granularity can serve as a better metric for black-box TSM. However, to the best of our knowledge, its impact has not yet been explored. 

\begin{figure*}
    \centering
    \includegraphics[width=1.05\linewidth]{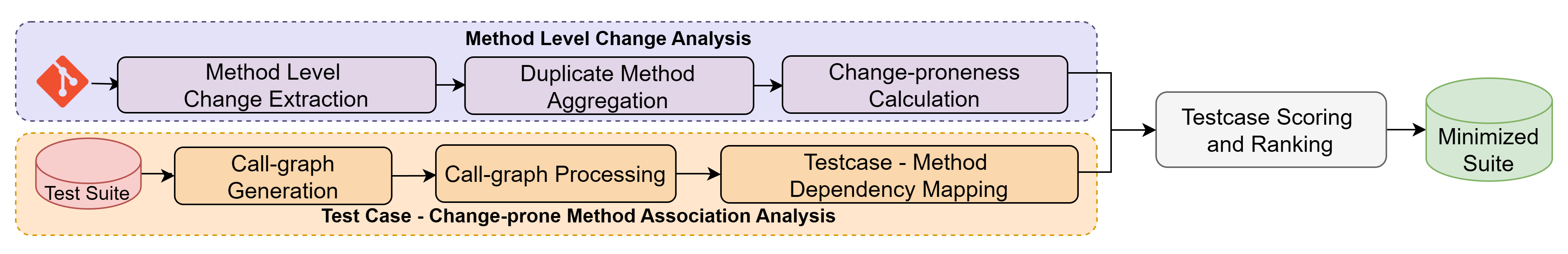}
    
    \caption{Approach Overview}
    \label{fig:overview}
\end{figure*}

\section{Methodology}

In this study, we evaluate the efficacy of CP as a metric for TSM by enhancing its granularity to method-level. Our goal is to investigate whether finer-grained change insights lead to more precise and effective minimization outcomes. Figure \ref{fig:overview} illustrates the methodology of our approach, \textit{MCTM}. 

\textit{MCTM} begins by mining method-level change histories from Git and computes CP scores based on the frequency and extent of changes. Next, a call graph is generated to uncover method-level dependencies for each test case. These dependencies are then used to score and rank test cases based on their association with change-prone methods. Finally, test suites are minimized under different budget levels to assess the impact of this finer-grained analysis.

\subsection{Method Level Change Information Extraction}

Changes in code are strongly associated with fault introduction, as frequent modifications often indicate instability and increased complexity in the affected components \cite{DBLP:conf/wcre/KociF0A24, DBLP:journals/ese/Rodriguez-Perez20}. Identifying such frequently changed regions can help focus testing efforts on fault-prone areas and improve the effectiveness of test suite minimization. To capture this phenomenon more precisely, we analyze change histories at the method level, moving beyond coarser class or file-level granularity to gain finer insights into localized changes.

Changes in code are often linked to fault introduction, making frequently modified methods strong indicators of fault-prone areas \cite{DBLP:conf/wcre/KociF0A24, DBLP:journals/ese/Rodriguez-Perez20}. To capture this, we extract three method-level change metrics: (a) the \textit{number of commits} in which a method was modified, reflecting how often it has changed; (b) the \textit{total number of commits} since the method was added, which provides context about its lifetime and exposure to change; and (c) the \textit{number of lines added, modified, and deleted}, indicating the scale of modifications. These metrics, adapted from prior work \cite{siam2025exploratory}, have been widely applied in fault prediction studies \cite{DBLP:journals/jksucis/RhmannPAP20, rahman2011ownership} and in analyzing class-level change trends \cite{DBLP:conf/ease/ArvanitouACA17}. By applying them at the method level, we aim to achieve finer-grained insights into change-prone regions. 

To capture a holistic view of historical changes, we compute these metrics from the project’s inception to the current commit. At the end of this step, we obtain a list of all modified methods along with their corresponding change metrics, forming a detailed historical profile of changes across the commit range.



\subsection{Change Metrics Aggregation}

\begin{table*}[h!]
\centering
\renewcommand{\arraystretch}{1.3}  
\setlength{\tabcolsep}{7.5pt}       
\caption{Commit Statistics for ExtendedBufferedReader::read()}
\begin{tabular}{lccccr}
\toprule
\textbf{Method} & \textbf{Change Commits} & \textbf{Total Commits} & \textbf{Insertions} & \textbf{Deletions} & \textbf{Remarks}\\
\midrule
ExtendedBufferedReader::read(char [ ] buf , int off , int len) & 4 & 573 & 91 & 49 & Original \\
ExtendedBufferedReader::read(char [ ] buffer , int off , int len) & 2 & 21 & 32 & 33 & Renamed \\
\rowcolor{gray!15}
ExtendedBufferedReader::read(char [], int, int) & 6 & 573 & 123 & 82 & Aggregated \\
\bottomrule
\end{tabular}
\label{tab:csvparser_stats}
\end{table*}

Due to Java’s support for polymorphism, we treat methods with the same name but different parameters as distinct functions. Each method is uniquely identified by its name, parameter types, and parameter identifiers. As a result, certain refactoring actions such as parameter renaming, are interpreted as the creation of a new method rather than a modification of an existing one. For instance, in TABLE \ref{tab:csvparser_stats}, the method ExtendedBufferedReader::read appears twice due to the renaming of the parameter from buf to buffer. Although semantically identical, they are treated as separate methods with distinct histories.
To handle such cases, we aggregate the duplicate method entries in the following manner:
\begin{itemize}
    \item We normalize method signatures by removing parameter identifiers to match methods that differ only in variable identifiers (e.g., \texttt{buf} vs. \texttt{buffer}).
    \item We add the change metrics i.e. \textit{number of changes, insertions, deletions}, and \textit{modifications} to capture the complete change history across all versions, resulting in the aggregated entry shown in the grey row of TABLE~\ref{tab:csvparser_stats}.
    \item We take the maximum \textit{total commit count} among the duplicates to preserve the full commit span since the method's creation (TABLE~\ref{tab:csvparser_stats}, Aggregated).
\end{itemize}

\subsection{Change-proneness Calculation}

Change proneness (CP) refers to the likelihood that a software entity will undergo modifications in the future~\cite{DBLP:conf/indiaSE/KumarLGM19}. Since past changes often indicate areas likely to change again, prior change history serves as a strong predictor of future evolution~\cite{DBLP:conf/wcre/KociF0A24, DBLP:journals/tse/ArisholmBF04, DBLP:journals/tse/KoruT05}. While Siam et al.~\cite{siam2025exploratory} applied CP at the class level, we advance this idea to the method level to capture localized change patterns with greater precision. This shift allows us to identify fault-prone regions that class-level analysis might miss.

We define change frequency (ChgFreq) as the proportion of commits where a method was modified relative to the total number of commits since its introduction (Eq.~\ref{eq:freq}). This metric reflects how often a method experiences modifications over its lifetime, highlighting areas of code that exhibit recurrent instability. 

\begin{equation} \label{eq:freq}
\text{ChgFreq}_{\mathit{M}} = \frac{\text{Number of Changes}_{\mathit{M}}}{\text{Total Commits}_{\mathit{M}}}
\end{equation}
Here, $M$ denotes the method under consideration.

To capture the extent of changes, we first compute \textit{code churn}, which is the total number of lines added, deleted, and modified (Eq.~\ref{eq:churn}). We then define \textit{change extent} (ChgExt) as the ratio of code churn to total commits, as shown in Eq.~\ref{eq:ext}. This formulation reflects how substantial the modifications are in each commit on average, distinguishing methods that undergo minor edits from those experiencing large, disruptive changes.

\begin{equation} \label{eq:churn}
\text{CodeChurn}_M = \text{Add}_M + \text{Del}_M + \text{Mod}_M
\end{equation}

\begin{equation} \label{eq:ext}
\text{ChgExt}_M = \frac{\text{CodeChurn}_M}{\text{Total Commits}_M}
\end{equation}

\subsection{Call-graph Generation}
A call graph is a directed graph that represents method invocations within a program. Each node corresponds to a method, and each edge indicates a call from one method to another \cite{DBLP:journals/ese/KeshaniGP24}. It highlights the dependencies between methods and helps in understanding how changes in one part of the code may affect others.

\begin{figure}
    \centering
    \includegraphics[width=1.15\linewidth]{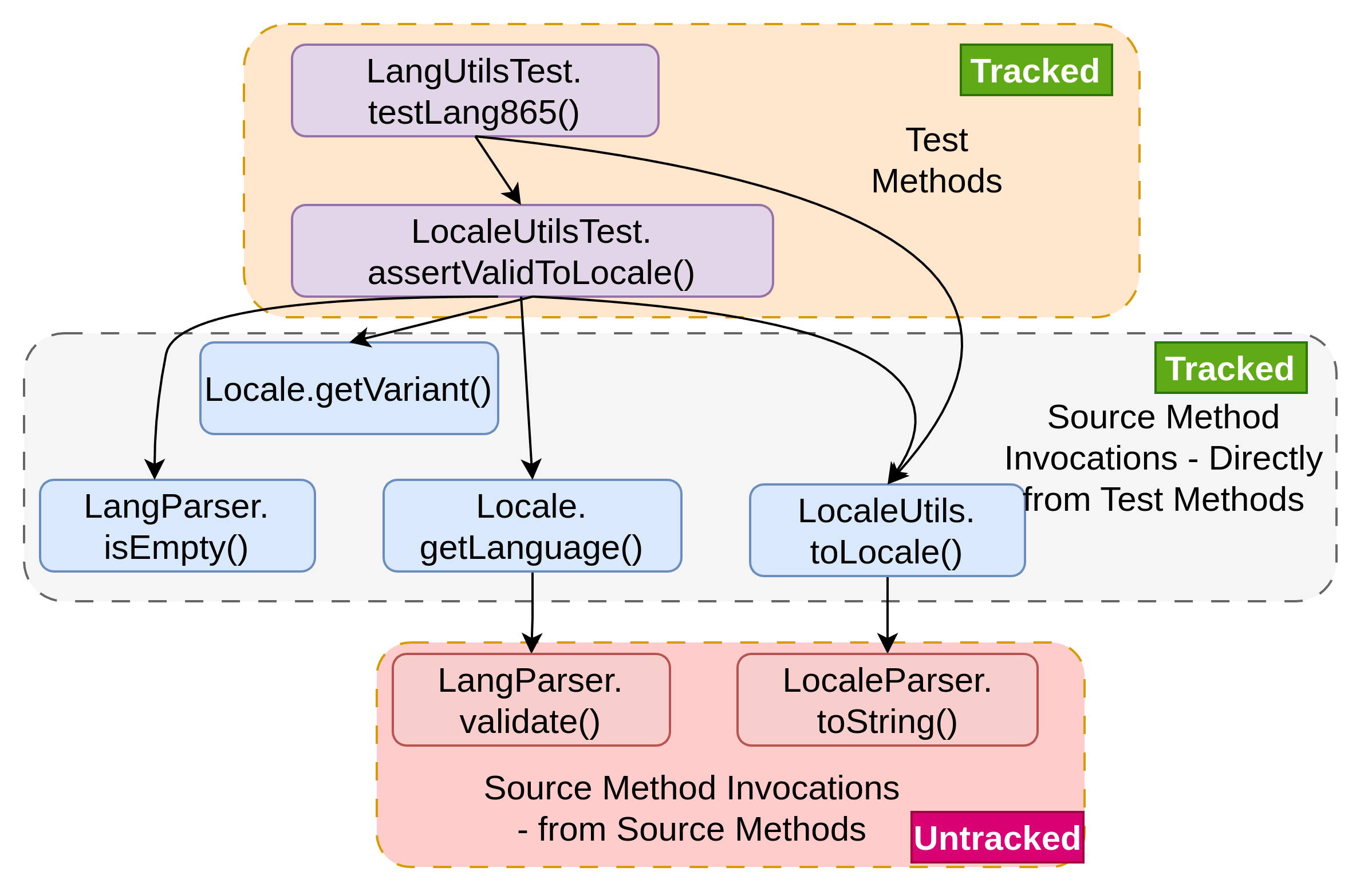}
    \hspace{-20em}
    \caption{Test case - Method Dependency Mapping Example}
    \label{fig:enter-label}
\end{figure}

For our analysis, we generate call graphs from the test code (orange-marked portion of Figure~\ref{fig:enter-label}) in each project to capture method-level dependencies. The test code is first compiled into JAR files, which serve as input for constructing the call graphs. We opt for static call graphs because they are efficient and suitable for a black-box setup. Unlike dynamic call graphs, which require executing the program and monitoring runtime behavior, static call graphs can be derived directly from compiled bytecode. This avoids the need for analyzing source code or system internals, aligning with our goal of minimizing test suites in a production-agnostic manner.

To ensure compatibility with our representation of changed methods, we further normalize the raw call graphs. This involves (1) replacing calls to \texttt{<init>} with their respective constructor names and (2) reformatting inner class method invocations by substituting \texttt{'\$'} with \texttt{'::'}. These steps produce fully processed call graphs, ready for method-level dependency analysis across all subject projects.

\subsection{Test case - Method Dependency Mapping}

In this step, we analyze the processed call graph to identify the method dependencies of each test case. Starting from each test method as the root node, we apply a depth-first search (DFS) to collect all directly and transitively invoked methods. This captures both immediate and hierarchical dependencies. 

For example, as shown in Figure \ref{fig:enter-label}, method \textit{LangUtilsTest::testLang865()} demonstrates multiple levels of method invocation in a hierarchical manner. The test method directly invokes \textit{LocaleUtilsTest::assertValidToLocale()} and \textit{LocaleUtils::toLocale()}. Subsequently, methods \textit{Locale::getVariant(), LangParser::isEmpty()} and \textit{Locale::getLanguage()} are invoked through \textit{LocaleUtilsTest::assertValidToLocale()}. Therefore we get a dependency list as \{\textit{LocaleUtilsTest::assertValidToLocale(), LocaleUtils::toLocale(), Locale::getVariant(), Locale::isEmpty(), Locale::getLanguage()}\}.

We also include the test method itself in the dependency list. This accounts for recently added tests that may still detect faults even if their dependent methods remain unchanged \cite{DBLP:journals/tse/HaasNJA24}. As a result, we obtain a multi-level hierarchical mapping of each test case to its associated method dependencies. Thus the final dependency list we get becomes \{\textit{LocaleUtilsTest::testLang865(), LocaleUtilsTest::assertValidToLocale(), LocaleUtils::toLocale(), Locale::getVariant(), Locale::isEmpty(), Locale::getLanguage()}\}.

Since we only packaged the test code into JARs, our call graphs capture only the method invocations originating from test regions (highlighted in orange and grey in Figure ~\ref{fig:enter-label}). However, subsequent hierarchical calls into production code (highlighted in red in Figure\ref{fig:enter-label}) are not captured, as these would require source code access.

\subsection{Test case Scoring and Ranking} \label{subsec:scoring}

A test case can depend on multiple methods, each with its own change-proneness (CP) value. To score test cases based on their association with change-prone methods, we extend the statistical measures introduced by \cite{siam2025exploratory} from class-level to method-level granularity. Originally, eight statistical measures: minimum, maximum, sum, mean, geometric mean, harmonic mean, standard deviation, and median—were proposed to summarize CP distributions. However, measures such as minimum and maximum are highly sensitive to outliers, which can distort test case prioritization by giving disproportionate weight to extreme CP values \cite{ahmadi2025robust, rousseeuw2011robust}. Similarly, standard deviation reflects the spread of CP values but can amplify the influence of outliers and is less informative for skewed distributions \cite{rousseeuw2011robust}. Moreover, using the sum as an aggregation metric can bias the score in favor of test cases with a large number of dependencies, potentially overshadowing test cases with a small number of dependencies in highly fault-prone areas. To address these concerns and achieve robust scoring, we considered only four statistical measures: arithmetic mean (Avg), geometric mean (GMean), harmonic mean (HMean), and median. These metrics balance sensitivity to variation with robustness to outliers and skewness, offering a more stable representation of method-level CP distributions for guiding test case selection.

Finally, we select test cases with the highest association scores under each measure. To reflect realistic developer expectations, we apply test suite minimization under three recommended budgets: 25\%, 50\%, and 75\% \cite{DBLP:conf/icse/PanGB23, siam2025exploratory, 10697930}.

\section{Experimental Setup and Result Analysis}
In this section, we evaluate \textit{MCTM} through a series of experiments. We begin by outlining the research questions that guide our study, followed by a description of the experimental setup, including the subject systems and evaluation metrics. We then present and analyze the results obtained from applying \textit{MCTM} to the test suites.

\subsection{Research Questions}

\textbf{RQ1: How effective is \textit{MCTM} to detect faults under different configurations?}

\textit{MCTM} utilizes 8 configurations, combining two CP measures (\textit{ChgFrq} and \textit{ChgExt}) and four statistical measures (Subsection \ref{subsec:scoring}) used to score a test case’s association with change-prone methods. This question explores how different configurations influence minimization effectiveness under three budget levels: 25\%, 50\%, and 75\%.

\textbf{RQ2: How does method-level change-proneness perform compared to class-level in terms of effectiveness and efficiency?}

In this question, we compare \textit{MCTM} (method-level) with \textit{CTM} (class-level) to assess whether the finer-grained method-level analysis in MCTM leads to improved fault detection while remaining efficient enough to be practical for black-box.

\textbf{RQ3: How does \textit{MCTM} compare with state-of-the-art TSM techniques that use other metrics?}

This question investigates whether method-level change-proneness can serve as a strong alternative to similarity-based methods. We evaluate \textit{MCTM} against two state-of-the-art black-box approaches: \textit{ATM}~\cite{DBLP:conf/icse/PanGB23}, which reduces test suite based on syntactic similarity, and \textit{LTM}~\cite{10697930}, which leverages large language models to detect and remove redundant tests. 

\begin{table}[htbp]
\centering
\caption{Subject Projects Statistics}
\label{tab:project_stats}
\footnotesize
\begin{tabular}{l@{\hspace{6pt}}r@{\hspace{6pt}}r@{\hspace{6pt}}r@{\hspace{6pt}}r@{\hspace{6pt}}r}
\toprule
\textbf{Project} & \textbf{Project} & \textbf{\# of} & \textbf{Tests} & \textbf{Avg \# of} & \textbf{Avg} \\
 & \textbf{Size} & \textbf{versions} & \textbf{Size} & \textbf{test cases} & \textbf{Commits} \\
 & \textbf{(KLoC)} & \textbf{(faults)} & \textbf{(KLoC)} & \textbf{per version} & \textbf{per version} \\
\midrule
Cli & 2 & 39 & 4 & 256 & 486 \\
Codec & 9 & 18 & 15 & 413 & 928 \\
Compress & 45 & 47 & 29 & 404 & 1,393 \\
Csv & 2 & 16 & 7 & 193 & 828 \\
Gson & 9 & 18 & 20 & 984 & 1,226 \\
JacksonCore & 31 & 26 & 45 & 356 & 902 \\
JacksonDatabind & 74 & 112 & 72 & 1,814 & 3,009 \\
JacksonXml & 6 & 6 & 10 & 152 & 660 \\
Jsoup & 14 & 93 & 13 & 494 & 785 \\
JxPath & 20 & 22 & 6 & 250 & 374 \\
Lang & 30 & 64 & 61 & 1,796 & 2,398 \\
Math & 71 & 106 & 73 & 2,078 & 2,932 \\
Mockito & 21 & 38 & 36 & 1,182 & 1,880 \\
Time & 30 & 26 & 56 & 3,918 & 1,600 \\
\bottomrule
\end{tabular}
\end{table}

\subsection{Experimental Setup}

We evaluated the performance of \textit{MCTM} under the eight configurations outlined in RQ1. All experiments were conducted on a machine with an Intel Core i5 9400F processor (6 cores, 2.9 GHz), 16 GB RAM, running Ubuntu 20.04. Since \textit{LTM}~\cite{10697930} relies on GPU-based inference, we equipped the machine with an additional GTX 1050 Ti GPU to support its evaluation. This represents an important distinction in computational requirements, as \textit{MCTM}, \textit{CTM}, and \textit{ATM} do not require GPU support.

Both \textit{MCTM} and \textit{CTM} produce deterministic results. In contrast, \textit{ATM} and \textit{LTM} involve stochastic processes due to their use of genetic algorithms (GA) and NSGA-II, resulting in non-deterministic outputs. To account for this variability, each approach was executed 10 times, and the reported results represent the average over these runs. All experiments were performed on the same machine to ensure consistency.

\subsubsection{Configurations}

A configuration refers to a specific combination of metrics and scoring methods used within an approach to prioritize and select test cases. Each technique we evaluated supports multiple such configurations, reflecting alternative ways of leveraging its underlying mechanisms.

For \textit{MCTM}, we explored eight configurations derived from two change metrics combined with four statistical aggregation measures (2x4=8). \textit{CTM}, which works at the class level, includes 16 configurations based on two CP measures and eight statistical strategies (2x8=16). \textit{ATM} offers six configurations, with four using similarity measurements paired with a genetic algorithm (GA) and two others employing NSGA-II. Likewise, \textit{LTM} provides 10 configurations by combining five large language models with two similarity measurements. To ensure a fair evaluation, we identified the best-performing configuration from each approach. This enables us to compare their performance effectively and gain insights into how different design choices influence fault detection capability.

\subsubsection{Minimization Budgets}
The minimization budget defines the proportion of test cases to retain after reduction. It represents the target size of the minimized test suite relative to the original suite. Similar to \cite{DBLP:conf/icse/PanGB23, 10697930}, we evaluated our approach under three budgets: 25\%, 50\%, and 75\%. These levels reflect commonly adopted practices in industry, where varying degrees of test suite reduction are applied to balance between fault detection capability and testing cost\cite{DBLP:conf/icse/PanGB23}.

\subsection{Dataset}
We evaluated our approach using projects from Defects4J\footnote{\url{https://github.com/rjust/defects4j}} (v2.0.1), a widely adopted benchmark dataset in software testing research~\cite{just2014defects4j}. Defects4J provides real and reproducible faults from open-source Java projects, along with the necessary test suites and fault-fixing commits. It is well-suited for test suite minimization (TSM) experiments, where linking test failures to system faults is essential\cite{DBLP:conf/icse/PanGB23, 10697930}.

Our dataset initially included all 16 projects from Defects4J v2.0.1, totaling 661 buggy versions. However, we excluded the Chart project due to unresolvable Git commit hashes that prevented successful metric extraction. This refinement resulted in 15 projects comprising 635 buggy versions for our experiments. Each buggy version contains a single real fault that triggers failures in one or more test cases.

Table~\ref{tab:project_stats} summarizes the key characteristics of the projects. The system sizes range from 2 KLoC to 74 KLoC, while the corresponding test suite sizes span 4 KLoC to 73 KLoC. The number of buggy versions per project varies between 4 and 174, with test suites containing an average of 152 to 3,919 test cases per version. These figures were determined by analyzing the latest version of each project using the CLOC tool.\footnote{\url{https://github.com/AlDanial/cloc}}. By utilizing this diverse set of projects, we ensure a comprehensive evaluation across small to large systems, enabling robust assessment of both effectiveness and scalability.

\subsection{Implementation Details}

Similar to prior studies~\cite{siam2025exploratory, 10697930}, we implemented our approach in Python to ensure portability and ease of integration with existing analysis tools. Several components of our method build upon established libraries and utilities. To extract method-level change metrics, we used PyDriller\footnote{\url{https://github.com/ishepard/pydriller/}}, a lightweight Python framework designed for mining software repositories\cite{spadini2018pydriller}. PyDriller provides a convenient API to access commit histories, analyze code churn, and identify fine-grained modifications, making it well-suited for processing version control metadata.

Although our approach is language-agnostic by design, this study focuses on Java projects. For call graph generation, we employed the java-callgraph\footnote{\url{https://github.com/gousiosg/java-callgraph}} tool, a lightweight utility capable of producing both static and dynamic call graphs from Java bytecode\cite{DBLP:conf/icsoft/JaszSPSF19}. Given the constraints of our black-box setup where production code and runtime execution are unavailable, we opted for static call graphs. This enabled us to efficiently capture method-level dependencies within test code without requiring program execution.


\subsection{Evaluation Metrics}

To evaluate the performance of \textit{MCTM}, we considered three metrics: \textbf{Accuracy}, \textbf{Fault Detection Rate (FDR)}, and \textbf{Execution Time}.

\textbf{Accuracy.} The primary goal of test suite minimization is to reduce the number of test cases while preserving fault detection capability. Accuracy measures how well this capability is maintained after minimization. Formally, as defined in Equation~\eqref{eq:accuracy}, accuracy is calculated as:

    \begin{equation}
        \text{Accuracy} = \frac{|F'|}{|F|}
        \label{eq:accuracy}
    \end{equation}
    
where $|F'|$ denotes the number of fault-revealing test cases retained in the minimized test suite, and $|F|$ is the total number of fault-revealing test cases in the original suite.

\textbf{Fault Detection Rate (FDR).} While accuracy reflects the proportion of fault-revealing tests retained, FDR provides a more relaxed perspective by focusing on whether any fault is detected after minimization. Specifically, FDR measures the percentage of buggy program versions for which the minimized test suite contains at least one fault-revealing test case\cite{DBLP:conf/icse/PanGB23, 10697930}.
Equation~\eqref{eq:fdr} defines FDR as:

    \begin{equation}
        \text{FDR} = \frac{\sum_{i=1}^{n} f_i}{n}
        \label{eq:fdr}
    \end{equation}
    
where $n$ is the total number of buggy versions, and $f_i = 1$ if the minimized suite for version $i$ includes at least one fault-revealing test case, and $f_i = 0$ otherwise. 

For example, in the \texttt{Codec} project, which has 18 faulty versions, if the minimized test suites detect faults in 15 versions but fail in 3, the FDR is computed as 15/18, yielding an FDR of 0.83.

\textbf{Execution Time.}  
To assess the efficiency of \textit{MCTM}, we measure the total execution time, which includes the time required for change metric extraction, call-graph generation, association analysis and scoring. This time is then compared with the total time required for \textit{CTM, ATM} and \textit{LTM}. 

\textbf{Fisher's exact test}
We used Fisher’s exact test \cite{raymond1995exact} to assess whether the differences in fault detection proportions between \textit{MCTM, CTM, ATM}, and \textit{LTM} are statistically significant. As a non-parametric test, it does not assume any underlying distribution and works well with small data samples, making it suitable for our data. A p-value below 0.05 indicates a significant difference in performance between the configurations.

\textbf{Odds ratio}
We computed the odds ratio \cite{arcuri2014hitchhiker} to quantify the strength of differences in fault detection between configurations. The odds ratio indicates how much more likely one configuration is to outperform another in terms of detected faults. An odds ratio greater than 1 suggests higher odds of better performance, while a value below 1 indicates the opposite. This measure complements Fisher’s exact test by providing an effect size for the observed differences.

\begin{table*}[htbp]
\centering
    \caption{Results and descriptive statistics of Accuracy of \textit{MCTM}, \textit{CTM}, \textit{ATM} and \textit{LTM} across projects for the 50\% minimization budget. The highest Accuracy results are highlighted in bold.}
    \setlength{\tabcolsep}{3.8pt}
    \renewcommand{\arraystretch}{1.3}
    \begin{tabular*}{\textwidth}{p{2cm}|*{7}{c|}c|c|c|c}
    \cline{1-12}
    \multirow{3}{*}{\diagbox[width=8.1em, height=4.3em]{Project}{Technique}} & \multicolumn{8}{c|}{\textbf{MCTM}} & {\textbf{CTM}} & \textbf{ATM} & \textbf{LTM} \\
    \cline{2-12}
    & \multicolumn{4}{c|}{\textbf{Change Frequency}} & \multicolumn{4}{c|}{\textbf{Change Extent}} & \textbf{Change Extent} & \textbf{Tree Edit Distance} & \textbf{Cosine Similarity} \\
    \cline{2-12}
    & Avg & HMean & GMean & Median & Avg & HMean & GMean & Median & GMean & Genetic Algorithm & UnixCoder \\
    \cline{1-12}
    Time            & \textbf{1.00} & 0.90 & \textbf{1.00} & 0.84 & \textbf{1.00} & 0.73 & \textbf{1.00} & 0.75 & 0.89 & 0.68 & 0.69 \\
    JxPath          & \textbf{0.77} & \textbf{0.77} & \textbf{0.77} & 0.73 & 0.75 & 0.73 & \textbf{0.77} & 0.68 & 0.55 & 0.67 & 0.75 \\
    Codec           & \textbf{0.84}          & 0.81 & \textbf{0.84}          & 0.83 & \textbf{0.84} & 0.76 & 0.83          & 0.77 & \textbf{0.84} & 0.65 & 0.65 \\
    JacksonXml      & \textbf{1.00} & 0.98 & 0.96 & 0.97 & \textbf{1.00} & 0.96 & 0.95 & 0.98 & 0.89 & 0.37 & 0.59 \\
    JacksonDatabind & \textbf{0.97}          & 0.86 & 0.96          & 0.82 & 0.96 & 0.72 & 0.94          & 0.80 & 0.68 & 0.63 & 0.77 \\
    Gson            & \textbf{1.00} & 0.83 & \textbf{1.00} & 0.89 & \textbf{1.00} & 0.81 & \textbf{1.00} & 0.83 & 0.73 & 0.66 & 0.68 \\
    Math            & 0.86          & 0.78 & 0.84          & 0.76 & \textbf{0.87} & 0.71 & 0.83          & 0.68 & 0.55 & 0.74 & 0.69 \\
    Jsoup           & \textbf{0.96}          & 0.81 & 0.96          & 0.73 & \textbf{0.96} & 0.79 & 0.97          & 0.71 & 0.53 & 0.61 & 0.71 \\
    Compress        & 0.91          & 0.76 & 0.89          & 0.68 & \textbf{0.94} & 0.73 & 0.91          & 0.65 & 0.79 & 0.84 & 0.76 \\
    Mockito         & \textbf{0.94} & 0.92 & 0.93          & 0.93 & 0.92 & 0.85 & 0.93          & 0.89 & 0.77 & 0.55 & 0.71 \\
    Cli             & \textbf{0.93} & 0.84 & 0.87          & 0.85 & 0.89 & 0.73 & 0.81          & 0.75 & 0.80 & 0.74 & 0.73 \\
    Csv             & \textbf{1.00} & 0.85 & 0.92          & 0.85 & 0.94 & 0.71 & 0.88          & 0.58 & 0.72 & 0.91 & 0.83 \\
    JacksonCore     & \textbf{0.99} & 0.76 & 0.95          & 0.72 & \textbf{0.99} & 0.75 & 0.87          & 0.67 & 0.86 & 0.53 & 0.53 \\
    Lang            & \textbf{0.94} & 0.91 & 0.90          & 0.81 & \textbf{0.94} & 0.78 & 0.92          & 0.84 & 0.72 & 0.72 & 0.70 \\
    \cline{1-12}
    \centering \textbf{Statistics} &       &       &       &       &       &       &       &       &       &       & \\
    Min            & \textbf{0.77} & 0.76  & \textbf{0.77} & 0.68 & 0.75 & 0.69 & \textbf{0.77} & 0.56 & 0.43 & 0.37 & 0.53 \\
    25\% Quartile  & \textbf{0.91} & 0.79  & 0.87          & 0.77 & 0.89 & 0.73 & 0.85          & 0.68 & 0.58 & 0.62 & 0.65 \\
    Mean           & \textbf{0.93} & 0.85  & 0.92          & 0.82 & \textbf{0.93} & 0.77 & 0.90          & 0.76 & 0.70 & 0.67 & 0.71 \\
    Median         & \textbf{0.95} & 0.84  & 0.93          & 0.83 & 0.94 & 0.74 & 0.92          & 0.74 & 0.73 & 0.66 & 0.70 \\
    75\% Quartile  & \textbf{0.99} & 0.90  & 0.96          & 0.85 & 0.98 & 0.79 & 0.96          & 0.83 & 0.80 & 0.74 & 0.74 \\
    Max            & \textbf{1.00} & 0.98 & 0.96 & 0.97 & \textbf{1.00} & 0.96 & 0.95 & 0.98 & 0.89 & 0.91 & 0.83 \\
    \cline{1-12}
\end{tabular*}
\label{tab:project-wise-accuracy}
\end{table*}

\begin{table*}[htbp]
\centering
    \caption{Results and descriptive statistics of FDR of \textit{MCTM}, \textit{CTM}, \textit{ATM} and \textit{LTM} across projects for the 50\% minimization budget. The highest FDR results are highlighted in bold.}
    \setlength{\tabcolsep}{3.8pt}
    \renewcommand{\arraystretch}{1.3}
    \begin{tabular*}{\textwidth}{p{2cm}|*{7}{c|}c|c|c|c}
    \cline{1-12}
    \multirow{3}{*}{\diagbox[width=8.1em, height=4.3em]{Project}{Technique}} & \multicolumn{8}{c|}{\textbf{MCTM}} & {\textbf{CTM}} & \textbf{ATM} & \textbf{LTM} \\
    \cline{2-12}
    & \multicolumn{4}{c|}{\textbf{Change Frequency}} & \multicolumn{4}{c|}{\textbf{Change Extent}} & \textbf{Change Extent} & \textbf{Tree Edit Distance} & \textbf{Cosine Similarity} \\
    \cline{2-12}
    & Avg & HMean & GMean & Median & Avg & HMean & GMean & Median & GMean & Genetic Algorithm & UnixCoder \\
    \cline{1-12}
    Time            & \textbf{1.00} & 0.96 & \textbf{1.00} & 0.84 & \textbf{1.00} & 0.76 & \textbf{1.00} & 0.80 & 0.72 & 0.92 & 0.80 \\
    JxPath          & 0.77 & 0.77 & 0.77 & 0.73 & 0.77 & 0.73 & 0.77 & 0.68 & 0.59 & 0.86 & \textbf{0.95} \\
    Codec           & \textbf{0.88}          & 0.81 &\textbf{ 0.88}          & \textbf{0.88} & \textbf{0.88} & 0.81 & \textbf{0.88}          & 0.81 & \textbf{0.88} & 0.83 & 0.83 \\
    JacksonXml      & \textbf{1.00} & \textbf{1.00} & \textbf{1.00}          & \textbf{1.00} & \textbf{1.00} & \textbf{1.00} & \textbf{1.00}          & \textbf{1.00} & \textbf{1.00} & 0.50 & 0.83 \\
    JacksonDatabind & \textbf{0.97}          & 0.88 & 0.96          & 0.84 & 0.97 & 0.75 & 0.94          & 0.82 & 0.85 & 0.67 & 0.80 \\
    Gson            & \textbf{1.00} & 0.83 & \textbf{1.00} & 0.89 & \textbf{1.00} & 0.83 & \textbf{1.00} & 0.83 & 0.72 & 0.78 & 0.78 \\
    Math            & 0.88          & 0.80 & 0.86          & 0.80 & \textbf{0.89} & 0.75 & 0.88          & 0.72 & 0.59 & 0.84 & 0.80 \\
    Jsoup           & \textbf{0.99}          & 0.86 & 0.98          & 0.77 & 0.98 & 0.83 & 0.99          & 0.78 & 0.47 & 0.71 & 0.86 \\
    Compress        & 0.91          & 0.77 & 0.89          & 0.70 & \textbf{0.94} & 0.74 & 0.91          & 0.66 & 0.87 & \textbf{0.91} & 0.85 \\
    Mockito         & \textbf{0.97}          & \textbf{0.97} & \textbf{0.97}          & \textbf{0.97} & \textbf{0.97} & 0.95 & \textbf{0.97}          & \textbf{0.97} & 0.87 & 0.74 & 0.89 \\
    Cli             & \textbf{0.95}          & 0.89 & 0.89          & 0.89 & 0.92 & 0.76 & 0.84          & 0.79 & 0.84 & 0.90 & 0.85 \\
    Csv             & \textbf{1.00} & 0.94 & 0.94          & 0.88 & 0.94 & 0.69 & 0.88          & 0.56 & 0.75 & \textbf{1.00} & 0.94 \\
    JacksonCore     & 0.96 & 0.77 & 0.96 & 0.73 & \textbf{1.00} & 0.77 & 0.88          & 0.69 & 0.73 & 0.65 & 0.69 \\
    Lang            & 0.93          & 0.95 & 0.93          & 0.87 & \textbf{0.98} & 0.84 & 0.97 & 0.89 & 0.57 & 0.81 & 0.80 \\
    \cline{1-12}
    \centering \textbf{Statistics} &       &       &       &       &       &       &       &       &       &       & \\
    Min            & \textbf{0.77} & \textbf{0.77} & \textbf{0.77} & 0.70 & \textbf{0.77} & 0.69 & \textbf{0.77} & 0.56 & 0.47 & 0.50 & 0.69 \\
    25\% Quartile  & \textbf{0.92} & 0.80  & 0.89          & 0.78 & \textbf{0.92} & 0.75 & 0.88          & 0.70 & 0.62 & 0.72 & 0.80 \\
    Mean           & \textbf{0.95} & 0.87  & 0.93          & 0.84 & 0.94 & 0.80 & 0.92          & 0.79 & 0.75 & 0.80 & 0.83 \\
    Median         & \textbf{0.97} & 0.87  & 0.95          & 0.86 & \textbf{0.97} & 0.77 & 0.93          & 0.79 & 0.74 & 0.83 & 0.82 \\
    75\% Quartile  & \textbf{1.00} & 0.95  & 0.98          & 0.89 & \textbf{1.00} & 0.83 & 0.99 & 0.83 & 0.86 & 0.88 & 0.85 \\
    Max            & \textbf{1.00} & \textbf{1.00} & \textbf{1.00} & \textbf{1.00} & \textbf{1.00} & \textbf{1.00} & \textbf{1.00} & \textbf{1.00} & \textbf{1.00} & \textbf{1.00} & 0.95 \\
    \cline{1-12}
\end{tabular*}
\label{tab:project-wise-fdr}
\end{table*}

\begin{table}[htbp]
\centering
    \caption{Time Comparison (in minutes) of Best Performing Configurations of MCTM, CTM, ATM and LTM}
    \setlength{\tabcolsep}{8pt}
    \renewcommand{\arraystretch}{1.3}
    \begin{tabular*}{\textwidth}{p{1.7cm}|*{3}{c|}c}
    \cline{1-5}
    \multirow{2}{*}{\diagbox[width=8.1em, height=4.3em]{Project}{Method}} & \textbf{MCTM} & \textbf{CTM} & \textbf{ATM} & \textbf{LTM} \\
    \cline{2-5}
    & \textbf{ChgFrq} & \textbf{ChgExt} & \textbf{Tree Edit} & \textbf{CosineSim.} \\
    \cline{2-5}
    & \textbf{Avg} & \textbf{GMean} & \textbf{Genetic A.} & \textbf{UnixCoder} \\
    \cline{1-5}
    Cli             & 0.17 & \textbf{0.14} & 1.06 & 0.43 \\
    Codec           & 0.49 & \textbf{0.30} & 2.74 & 0.41 \\
    Compress        & 0.82 & \textbf{0.45} & 3.55 & 0.58 \\
    Csv             & 0.31 & \textbf{0.26} & 0.63 & 0.26 \\
    Gson            & 0.64 & \textbf{0.36} & 17.05 & 1.48 \\
    JacksonCore     & 0.63 & \textbf{0.27} & 2.18 & 0.47 \\
    JacksonDatabind & 2.66 & \textbf{0.90} & 46.97 & 3.26 \\
    JacksonXml      & 0.17 & 0.19 & 0.39 & \textbf{0.16} \\
    Jsoup           & 0.42 & \textbf{0.22} & 4.68 & 0.44 \\
    JxPath          & 0.36 & \textbf{0.13} & 0.93 & 0.24 \\
    Lang            & 1.90 & \textbf{0.76} & 66.25 & 5.12 \\
    Math            & 2.54 & \textbf{0.99} & 140.73 & 12.75 \\
    Mockito         & 0.89 & \textbf{0.68} & 33.68 & 1.67 \\
    Time            & 1.65 & \textbf{0.52} & 320.40 & 7.97 \\
    \cline{1-5}
    \centering \textbf{Statistics} &       &       &       &       \\
    Min            & 0.17 & \textbf{0.13} & 0.39 & 0.16 \\
    25\% Quartile  & 0.38 & \textbf{0.23} & 1.34 & 0.38 \\
    Mean           & 0.98 & \textbf{0.44} & 45.80 & 2.60 \\
    Median         & 0.63 & \textbf{0.33} & 4.12 & 0.53 \\
    75\% Quartile  & 1.46 & \textbf{0.64} & 43.65 & 2.93 \\
    Max            & 2.66 & \textbf{0.99} & 320.40 & 12.75 \\
    \cline{1-5}
\end{tabular*}
\label{tab:time_comparison_adapted}
\end{table}

\subsection{Results}

We applied our methodology using three minimization budgets: 25\%, 50\%, and 75\%. Consistent with prior studies~\cite{siam2025exploratory, 10697930, DBLP:conf/icse/PanGB23}, we center our analysis on the 50\% budget, as results across the other budgets (25\% and 75\%) exhibited similar trends and led to the same conclusions. The data, codes and results of our experiments are available in our replication package\cite{figshare_data_2025}. 

\subsubsection{\textbf{RQ1 Results}}

Table~\ref{tab:project-wise-accuracy}, \ref{tab:project-wise-fdr}, and \ref{tab:time_comparison_adapted} present the project-wise accuracy, fault detection rate (FDR), and execution time, respectively, of the eight \textit{MCTM} configurations under a 50\% minimization budget. Bold values highlight the best-performing configurations for each statistic. All configurations share the same underlying method-level change metric extraction, call graph generation, and dependency analysis. Since the time differences among configurations arise only from scoring, which is computationally inexpensive, we report execution time of \textit{MCTM} as a single column. On average, \textit{MCTM} completed processing in 58.40 seconds per version, with a median of 38.16 seconds.

Among the eight configurations, the combination of method-level \textit{ChgFrq} with \textit{Average} aggregation consistently achieved the strongest results. It reached a mean accuracy of 0.93 and FDR of 0.94, ranking highest across most subject projects. This suggests that aggregating change frequencies evenly across all associated methods effectively captures their collective evolution patterns, enabling better prioritization of fault-revealing test cases. 

In contrast, using \textit{Median} aggregation produced the weakest results (mean accuracy: 0.82, mean FDR: 0.84). This likely stems from the \textit{Median} masking extreme values in the associated methods, which are critical for identifying highly change-prone components. By scoring test cases based on a middle value, it underrepresents cases where a few highly volatile methods dominate a test’s behavior, leading to suboptimal prioritization.

Configurations employing \textit{ChgExt} exhibited slightly lower performance compared to \textit{ChgFrq}. While \textit{ChgExt/Average} was the strongest within this group, it still was outperformed by \textit{ChgFrq/Average}. This may be attributed to \textit{ChgExt}’s broader numerical range, which amplifies variance during aggregation. Interestingly, aggregation methods like \textit{Geometric Mean}, which dampen outliers, showed more stable results across both metrics, highlighting their potential in situations where extreme values are common.

Fisher’s exact test further supports these findings. The best configuration (\textit{ChgFrq/Average}) significantly outperformed all other configurations except \textit{ChgExt/Average} at the \(\alpha < 0.05\) level. While \textit{ChgFrq/Average} achieved higher accuracy and FDR in seven projects (\texttt{JxPath, Codec, JacksonXml, JacksonDatabind, Mockito, Cli, Csv}), the observed differences against \textit{ChgExt/Average} were not statistically significant.

As observed in Table~\ref{tab:project-wise-accuracy} and \ref{tab:project-wise-fdr}, the lowest accuracy and FDR for the best \textit{MCTM} configuration (0.77) occurred in the \texttt{JXPath} project. A manual inspection revealed that this drop was due to limitations of static analysis in our black-box setup. Specifically, inherited methods in test code were not fully captured because static call graphs were generated only from compiled test code. Inheritance plays a key role in \texttt{JXPath}, leading to incomplete dependency mapping and reduced selection effectiveness. Despite this, the overall consistently high accuracy and FDR across all projects affirm the robustness of method-level change metrics for black-box test suite minimization.


\subsubsection{\textbf{RQ2 Results}}

We compare the effectiveness and efficiency of method-level (\textit{MCTM}) and class-level (\textit{CTM}) change-proneness. As shown in TABLE~\ref{tab:project-wise-accuracy} and \ref{tab:project-wise-fdr}, \textit{MCTM} consistently outperforms \textit{CTM} across all statistical measures. On average, \textit{MCTM} achieved 0.93 accuracy and 0.94 FDR, while \textit{CTM} reached only 0.70 and 0.75, respectively. Fisher’s exact test further confirms the significance of this improvement (\textit{p} = 0.004, Odd's Ratio = 23.4), indicating that \textit{MCTM} is over 23 times more likely to achieve superior fault detection. The difference is especially notable in project \texttt{Jsoup} (0.96 vs 0.53 accuracy), where classes often follow a \textit{Chain of Responsibility} pattern \cite{DBLP:books/daglib/0011977}. Here, method-level granularity uncovers fine-grained dependencies missed by static black-box class-level analysis \cite{siam2025exploratory}. These findings highlight the value of method-level metrics in guiding fault-revealing test selection.

In terms of efficiency (TABLE~\ref{tab:time_comparison_adapted}), \textit{MCTM} incurs a modest overhead, averaging 0.98 minutes per minimization run compared to 0.48 minutes for \textit{CTM}. This increase stems from the deeper method-level analysis but remains within practical limits. The highest observed time was 2.66 minutes for \texttt{JacksonDatabind}, the largest project in our dataset (74 KLoC, 3000+ commits, 1814 tests per version). Since test suite minimization is performed infrequently in real-world workflows \cite{DBLP:conf/swqd/NoemmerH20}, this additional cost is acceptable given the substantial improvements in fault detection.


\subsubsection{\textbf{RQ3 Results}}

To answer RQ3, we compared \textit{MCTM} with the best configurations of two state-of-the-art black-box TSM techniques: \textit{ATM} and \textit{LTM}. Table~\ref{tab:project-wise-accuracy}, \ref{tab:project-wise-fdr} and \ref{tab:time_comparison_adapted} summarizes their performance in terms of accuracy, FDR, and execution time respectively under a 50\% minimization budget.

\textit{MCTM} achieved a mean accuracy of 0.93 and FDR of 0.94, significantly outperforming all baseline methods. In comparison, \textit{ATM} and \textit{LTM} achieved 0.67/0.81 and 0.71/0.84 for accuracy and FDR, respectively. This substantial gain highlights the strength of method-level change-proneness in guiding test case selection. While similarity-based approaches like \textit{ATM} and \textit{LTM} aim to minimize overall redundancy, they often exclude fault-revealing tests when similar test cases exist. This inter-test dependency reduces their fault detection potential. Such case occurs in project \texttt{JacksonXml} (0.37 accuracy with \textit{ATM} and 0.50 with \textit{LTM}), where occurrence of test cases are prominent which are very similar to the bug revealing ones. In contrast, \textit{MCTM} evaluates each test case independently, scoring them based on their association with highly change-prone methods. This avoids such dependencies and ensures a higher likelihood of retaining fault-revealing tests. 

The superiority of \textit{MCTM} over \textit{ATM} and \textit{LTM} is statistically validated using Fisher’s exact test. Against \textit{ATM}, \textit{MCTM} achieved an odds ratio of 78.0 (\textit{p} = 0.000069). The comparison with \textit{LTM} yields an even stronger odds ratio of 169.0 (\textit{p} = 0.000010). Both results are highly significant ($p < 0.001$) and supported by tight confidence intervals, confirming the strength of these findings.
 
In terms of execution time, \textit{MCTM} averaged 0.98 minutes per version, compared to \textit{ATM}'s 45.80 minutes and \textit{LTM}'s 2.60 minutes. Even though \textit{LTM} benefits from GPU acceleration (GTX 1050 Ti), it is still 2.65x slower than \textit{MCTM}, which runs entirely on CPU. The high runtime of \textit{ATM} and \textit{LTM} arises from their computationally intensive all-pair similarity calculations and evolutionary search heuristics. In contrast, as discussed below in Subsection \ref{sub:time-complexity}, \textit{MCTM} employs a lightweight, per-test analysis of change-proneness and dependencies, resulting in a linear-time process that scales efficiently with project size.

These findings collectively demonstrate that method-level change-proneness is not only a more effective metric for black-box TSM but also enables highly efficient implementations suitable for large-scale systems.

\section{Discussion}

\subsection{Time Complexity} \label{sub:time-complexity}
The efficiency of our approach stems from its reliance on lightweight analysis of version control metadata. Both \textit{CTM} and \textit{MCTM} achieve a time complexity of \(O(n \cdot m)\), where \(n\) represents the total number of commits and \(m\) the number of modified methods or classes on a commit. This enables them to scale well even for projects with extensive commit histories and large codebases. In contrast, similarity-based techniques such as \textit{ATM} and \textit{LTM} involve computing pairwise similarities across all test cases (\(O(t^2 \cdot d_1 \cdot d_2)\), where \(t\) is the number of test cases and \(d_1, d_2\) are the AST depths of the respective test cases). These similarity computations are further coupled with evolutionary search algorithms (GA and NSGA-II), which operate over large populations and multiple generations. As a result, their overall complexity is exponential, leading to significantly higher execution times. This distinction is particularly relevant for large industrial systems where test suites may contain thousands of test cases, making scalability a critical factor.

\subsection{Accuracy and FDR Trends}  
A closer look at fault detection performance reveals interesting patterns across the different approaches. \textit{MCTM} demonstrates comparable accuracy and FDR (0.93 mean accuracy and 0.95 mean FDR for \textit{MCTM}). This similarity arises because both methods prioritize test cases associated with highly change-prone components. In the Defects4J dataset, multiple fault-revealing test cases often target the same change-prone regions. Consequently, once these regions are identified and prioritized, most relevant tests are naturally included in the reduced suite. This leads to high coverage of fault-detecting tests, resulting in stable accuracy and FDR across different minimization budgets.  

On the other hand, \textit{ATM} and \textit{LTM} focus on reducing redundancy by minimizing the overall similarity among selected test cases. While this strategy promotes diversity and increases the likelihood of capturing at least one fault-revealing test (thus boosting FDR), it can also deprioritize similar tests targeting the same fault. As a result, accuracy may suffer because not all fault-revealing test cases are retained. For example, TABLE \ref{tab:project-wise-accuracy} shows the mean accuracy of ATM is 0.67, while the mean FDR is 0.80 (TABLE \ref{tab:project-wise-fdr})  This trade-off between diversity and thoroughness is inherent in similarity-based approaches and highlights a key difference in their behavior compared to change-proneness-based techniques.

\subsection{Hardware Considerations} 
It is also important to consider the computational requirements of these techniques. \textit{LTM} relies on GPU-based inference to process large language model embeddings, requiring hardware support beyond standard CPU resources. In contrast, \textit{MCTM}, \textit{CTM}, and \textit{ATM} operate entirely on CPU, making them more accessible for environments with limited hardware. Despite \textit{LTM} benefiting from GPU acceleration (via a GTX 1050 Ti), our experiments show that \textit{MCTM} consistently achieves lower execution times. This highlights its superior efficiency even under less favorable computational conditions and reinforces its suitability for practical adoption in resource-constrained environments.

\subsection{Summary of Trade-offs}
The comparative analysis of \textit{MCTM}, \textit{CTM}, \textit{ATM}, and \textit{LTM} reveals distinct trade-offs between accuracy, fault detection robustness, computational efficiency, and hardware requirements. Change-proneness-based techniques (\textit{MCTM}, \textit{CTM}) excel in efficiency and scalability due to their lightweight reliance on version control metadata and static analysis. Their ability to prioritize tests covering highly change-prone regions leads to consistently high accuracy and fault detection rates, making them particularly effective for large projects with extensive test suites. However, this focus can potentially overlook tests unrelated to change-prone areas, which might be necessary for detecting less frequent, isolated faults.

Similarity-based methods (\textit{ATM}, \textit{LTM}) emphasize test diversity, improving the likelihood of detecting faults across varied code paths. While this strategy enhances fault detection robustness in heterogeneous test suites, it comes at the cost of higher execution times and, in the case of \textit{LTM}, dependency on GPU hardware. Their computational complexity limits scalability in large industrial systems where thousands of test cases exist.

Overall, \textit{MCTM} demonstrates a strong balance by achieving fine-grained fault localization through method-level CP analysis while retaining practical efficiency. This balance makes it a promising choice for real-world applications, where both effectiveness and scalability are critical.


\section{Threats to Validity}

\subsection{Construct Validity}
Construct validity examines whether the chosen metrics and methods accurately capture the effectiveness of TSM. We operationalize TSM effectiveness as the ability to retain tests that are likely to reveal faults while minimizing test suite size and execution overhead. Our central construct is that method-level change proneness serves as a valid proxy for fault proneness. 

Our approach relies on method-level change proneness metrics derived from version control metadata. While version control histories may not fully reflect semantic changes such as  refactoring or formatting changes that do not alter program behavior, they have been widely used as proxies for code evolution in prior fault prediction and TSM studies~\cite{DBLP:journals/ese/Rodriguez-Perez20, DBLP:journals/tse/KoruT05, siam2025exploratory}. We mitigate such noise by considering changes over the entire commit history, rather than relying on a small set of recent commits that may amplify transient fluctuations.

Additionally, we utilize static call graphs to analyze test case dependencies. While static analysis may miss certain dynamic behaviors (e.g., inheritance, polymorphism resolved at runtime), it offers a practical and scalable means to approximate method interactions without requiring program execution. This is particularly important in our black-box setup, where we intentionally avoid reliance on production code or runtime environments. The effectiveness of static call graphs in similar contexts has been demonstrated in prior work~\cite{luo2016large}, further supporting their use here.

\subsection{Internal Validity}
Internal validity relates to whether observed effects are genuinely due to our methods rather than confounding factors.

One notable concern is hardware disparity: \textit{LTM} requires GPU-based inference to support large language models, whereas \textit{MCTM, CTM,} and \textit{ATM} run entirely on CPU. This difference could bias efficiency comparisons in favor of \textit{LTM}. However, despite \textit{LTM} benefiting from GPU acceleration (GTX 1050 Ti), our results show that \textit{MCTM} consistently outperforms it in execution time, highlighting \textit{MCTM}’s superior efficiency even under less favorable hardware conditions.

We also recognize that \textit{ATM} and \textit{LTM} involve stochastic optimization techniques (GA and NSGA-II), which may produce non-deterministic results across runs. To address this variability, we executed each non-deterministic configuration 10 times and reported the average results. In contrast, \textit{MCTM} and \textit{CTM} are deterministic, ensuring stable and reproducible outputs. Additionally, all experiments were conducted on the same hardware platform to avoid variability introduced by differing system environments.


\subsection{External Validity}
External validity considers the generalizability of our findings beyond the studied dataset and context.

Our experiments are based on 15 Java projects from Defects4J v2.0.1, covering 635 buggy versions. While these projects are open-source and may differ from industrial systems, Defects4J remains the most widely adopted benchmark in TSM and fault prediction research due to its collection of real-world faults and reproducible test suites~\cite{just2014defects4j}. This ensures our findings are relevant to the broader software engineering community. We also note that Defects4J versions contain a single fault per buggy version, whereas real-world systems often exhibit multiple interacting faults. While this is a limitation of all studies using Defects4J, it does not diminish the value of our work, as single-fault scenarios provide a controlled environment for evaluating fault detection capability. Future work can extend our evaluation to multi-fault datasets as they become available.

Although our focus on Java projects may raise questions about applicability to other languages, the core principles of change-proneness and test code analysis are language-agnostic. Our approach can generalize to other ecosystems with minor adaptations.

\subsection{Conclusion Validity}
Conclusion validity pertains to whether the statistical analyses applied lead to accurate and trustworthy inferences.

We employed Fisher’s exact test, a non-parametric test, suited to small sample sizes and categorical data, to assess whether differences in fault detection rates are statistically significant. Additionally, we reported odds ratios to quantify the strength of these differences. While multiple pairwise comparisons can increase the risk of Type I errors, our consistent findings across configurations and projects provide strong evidence of the robustness of our results.

\section{Conclusion and Future Work}

This study investigates the use of method-level CP as a metric for black-box TSM and incorporates it into an approach, \textit{MCTM}. To select test cases for the reduced suite, \textit{MCTM} extracts method level change-metrics to calculate CP, analyzes test case's association with change-prone methods and scores test cases using statistical measures (average, median etc.). We evaluated \textit{MCTM} on a large dataset under 8 different configurations and identified the optimal \textit{MCTM} configuration (\textit{ChgFrq} with \textit{Average} aggregation). Comparative analysis with state-of-the-art TSM approaches demonstrates that \textit{MCTM} achieves substantially higher effectiveness with 0.93 accuracy and 0.94 FDR while remaining efficient and lightweight. The findings demonstrate that, as a metric, method-level CP can guide for effective black-box TSM. 

Future work can examine how results vary across different datasets, as well as the impact of statement-level CP and adaptation to other widely-used programming languages.

\bibliography{CP-TSM} 
\bibliographystyle{IEEEtran}

\end{document}